\documentstyle[preprint,aps,epsfig]{revtex}
\tighten
\begin{document}

\preprint{BIHEP-TH-2003-6}

\title{ $B_{s,d} \to \gamma\gamma$ Decay with the  Fourth Generation }
\author{ Wujun Huo$^{a}$, Cai-Dian  L\"{u}$^{b}$, and
 Zhenjun Xiao$^{c}$  \\
{\small\it a.  Department of  Physics, Peking University, Beijing
$100871$, P.R. China}
 \\
{\small\it b.  CCAST (World Laboratory), P.O.Box 8730, Beijing
              100080, P.R.China;} \\
{\small\it b. Institute of High Energy Physics, CAS, P.O. Box
918(4),
Beijing 100039, P.R.China\thanks{Mailing address}}\\
{\small\it c. Physics Department, Nanjing Normal University,
Nanjing, Jiangsu 210097, P.R.China}}
\date{\today}
\maketitle

\begin{abstract}
We calculate  the branching ratios of $B_{s,d}\to  \gamma \gamma$
decay in the sequential fourth generation model. We find that the
theoretical values of the branching ratios, ${\rm
BR}(B_{s}\to\gamma\gamma)$, including the contributions of $t'$,
are much different from the minimal standard model (SM)
predictions. The new physics contribution can provide one to two
orders enhancement to the standard model prediction of ${\rm
BR}(B_{s}\to\gamma\gamma)$. But due to the tiny values of 4th
generation CKM matrix element $V_{t'd}^* V_{t'b}$, the new physics
effect on ${\rm BR}(B_{d}\to\gamma\gamma)$  is very small and can
not be distinct from the SM prediction. It is shown that the decay
$B_s\to \gamma\gamma$ can give the test of new physics signals
from the 4th generation model.
\end{abstract}
\vspace{1cm}


\newpage
\section{Introduction}
The minimal Standard Model (SM) is a very successful theory of the
elementary particles known today. But it should not be the final
theory, because it has too many unknown parameters to be put by
hand. Most of these parameters are in the fermion part of the
theory. We don't know the origin of the quarks and leptons, as
well as how to determinate their mass and number theoretically. We
have to get their information all from experiments. There is still
not  a successful theory which can
 unify all the four basic interactions.
Perhaps our {\it elementary particles} have substructure
 and we need a more fundamental  theory than SM. This is beyond the scope of our
 current experiments.  On the other hand, the recent
measurement of the muon anomalous magnetic moment by the
experiment E821 \cite{e821} disagrees with the SM expectations at
more than 1.6$\sigma$ - 3$\sigma$ \cite{new1,new2} level.
Moreover, there are convincing evidences that neutrinos are
massive and oscillate in flavor \cite{neutrino}. It seems to
indicate the presence of new physics. The discovery of neutrino
oscillation has been one of the most exciting experimental results
in the recent years \cite{first}. As the experimental data
accumulating in the Super-Kamiokande collaboration \cite{new,sma},
  more results published by SNO \cite{sno}, K2K \cite{k2k} and
CHOOZ \cite{chooz} experiments. These experiments give possible
signals of  new physics beyond SM.

 In this paper, we consider the sequential fourth generation model to
 estimate the possible contributions
to the decay  $B_{s,d}\to \gamma\gamma$ from the exchange of the
fourth up-type quark $t'$. From the point of phenomenology,
 there is a realistic question one may ask: What is  the number of the
fermion generations or weather there are  additional quarks or
leptons other than the known three generations. The present
experiments can tell us there are only three generation of
fermions with $light$ neutrinos whose mass is smaller than $M_Z
/2$ \cite{Mark}, but the experiments don't exclude the existence
of other kind of additional generation, such as the fourth
generation, with a $heavy$ neutrino, i.e. $m_{\nu_4} \geq M_Z /2$
\cite{Berez}. Latest studies in the electroweak sector allow the
existence of a fourth generation with heavy Dirac neutrino
\cite{he}. Many authors studied models which extend the fermions
part, such as vector-like quark models \cite{vec-like}, sterile
neutrino models \cite{sterile} and the sequential four generation
standard model (SM4) \cite{McKay}. Here  we consider the
sequential fourth generation, non -supersymmetric model, which
includes an up-like quark $t'$, a down-like quark $b'$, a lepton
$\tau'$, and a heavy neutrino $\nu'$ besides the known three
generation of fermions
 in the SM. The properties of these new
fermions are  the same as their corresponding counterparts
of other three generations except their masses and CKM matrix elements.

Many attempts have been made recently about the fourth generation
\cite{fourth1,fourth2,fourth3} and the experimental searches of
the fourth generation particles \cite{exp1,exp}. In our previous
papers \cite{huo1,huo2,hwj}, we investigated the rare $B$ meson
decays with the fourth generation \cite{huo1}, $\epsilon^{'}
/\epsilon$ in $K^0$ systems  and $\Delta M_{B_{s,d}}$ in SM4
\cite{huo2}. We got some interesting results, such as the new
effects of the 4th generation particle on the meson decays and CP
violation. We also got the constraints of the fourth generation
CKM matrix factors \cite{huo1,huo2}.  Recently, these decays in
fourth generation were considered again. In ref. \cite{fourth1},
the phases of the fourth generation CKM matrix factors,
$V_{t^{'}s}^{*} V_{t^{'}b}$, have been included. The exclusive B
decays with the fourth generation were investigated in Ref.
\cite{fourth2}. These rare decays are very interesting to study
the new physics effects of the fourth generation. In this paper,
we would like to investigate the rare decay $B_{s,d} \to
\gamma\gamma$ with the fourth generation.

As a loop-induced flavor changing neutral current (FCNC) process
the inclusive  decay (at quark level) $b\to s \gamma \gamma$ is in
particular sensitive to contributions from those new physics
beyond the SM. There is a vast interest in this decay. On the
experimental side, only  upper limits ($90\%C.L.$) on the
branching ratios of $B_{s,d} \to \gamma \gamma$ decay are
currently available
\begin{eqnarray}
{\rm BR}(B_s\to \gamma \gamma )& <& 1.48 \times 10^{-4}, \cite{cleo95}, \\
{\rm BR}(B_d\to \gamma \gamma )& <& 1.7 \times 10^{-6},
\cite{babar01},
\end{eqnarray}
which are roughly two orders above the SM predictions
\cite{yao,hiller,bosch02},
\begin{eqnarray}
5 \times 10^{-7} <  {\rm BR}(B_s\to \gamma \gamma) <  15\times 10^{-7}, \\
1 \times 10^{-8} <  {\rm BR}(B_d\to \gamma \gamma) < 4.0 \times
10^{-8}.
\end{eqnarray}
Within the SM, the electroweak contributions to $b\to s
\gamma\gamma $ and $B \to \gamma\gamma$ decays have been
calculated some time ago \cite{yao}, the leading-order QCD
corrections and the long-distance contributions were evaluated
recently by several groups \cite{hiller,long}. The new physics
corrections were also considered, for example, in the two-Higgs
doublet model \cite{2hd,xiao01} and the supersymmetric model
\cite{susy}. There is also study concerning other exclusive
channels \cite{exc}. In a previous paper \cite{xiao03}, we
calculated the Technicolor corrections to the rare decays $B_{s,d}
\to \gamma \gamma$ in the QCD factorization approach. The great
progress in theoretical studies recently encourage us to do more
investigations about this decay in the sequential fourth
generation model.

This paper is organized as the following: In Section 2, we present
the formulae for  the decays $B_{s,d}\to \gamma\gamma$ in SM and
the fourth generation model. We also obtain the constraints of the
fourth generation CKM matrix factors, $V^*_{t'd}V_{t'b}$ and
$V^*_{t's}V_{t'b}$ from experiments in this section.   In Section
3, we give the numerical analysis of the branching ratios in the
fourth generation model. The conclusions are in the last section.

\section{ $B_{s,d}\to \gamma\gamma$ in the fourth generation model}

 Same as in our last paper \cite{xiao03}, up to the corrections of order
$1/m_W^2$,
  the effective Hamiltonian for
 $b\to s\gamma\gamma$ at scales   $\mu_b={\cal O}(m_b)$ is just the one for
 $b\rightarrow s \gamma$ and takes the form
    \begin{equation} \label{Heff_at_mu}
        {\cal H}_{\rm eff} =
           \frac{G_{\rm F}}{\sqrt{2}} V_{ts}^* V_{tb}
           \left[ \sum_{i=1}^6 C_i(\mu_b) Q_i +
           C_{7\gamma}(\mu_b) Q_{7\gamma}
          +C_{8G}(\mu_b) Q_{8G} \right]\,, \label{eq:heff}
    \end{equation}
 where in view of $|V^{*}_{us} V_{ub} /V^{*}_{ts} V_{tb} | <0.02$, we
 have neglected the term proportional to $V^{*}_{us} V_{ub}$. Here $Q_1 \dots
 Q_6$ are the usual four-fermion operators whose explicit form is given below.
  The last two operators in the Eq.(\ref{Heff_at_mu}),
 characteristic for this decay, are the {\it magnetic--penguin} operators.
 The complete list of operators is given as follows
\begin{eqnarray}
O_1&=&(\overline{c}_{L\beta} \gamma^{\mu} b_{L\alpha})
            (\overline{s}_{L\alpha} \gamma_{\mu} c_{L\beta})\;,\\
O_2&=&(\overline{c}_{L\alpha} \gamma^{\mu} b_{L\alpha})
            (\overline{s}_{L\beta} \gamma_{\mu} c_{L\beta})\;,\\
O_3&=&(\overline{s}_{L\alpha} \gamma^{\mu} b_{L\alpha})
\sum_{q=u,d,s,c,b}(\overline{q}_{L\beta} \gamma_{\mu} q_{L\beta})\;,\\
O_4&=&(\overline{s}_{L\alpha} \gamma^{\mu} b_{L\beta})
\sum_{q=u,d,s,c,b}(\overline{q}_{L\beta} \gamma_{\mu} q_{L\alpha})\;,\\
O_5&=&(\overline{s}_{L\alpha} \gamma^{\mu} b_{L\alpha})
\sum_{q=u,d,s,c,b}(\overline{q}_{R\beta} \gamma_{\mu} q_{R\beta})\;,\\
O_6&=&(\overline{s}_{L\alpha} \gamma^{\mu} b_{L\beta})
\sum_{q=u,d,s,c,b}(\overline{q}_{R\beta} \gamma_{\mu} q_{R\alpha})\;,\\
O_7&=&(e/16\pi^2) m_b \overline{s}_L \sigma^{\mu\nu}
            b_{R} F_{\mu\nu}\;,\\
O_8&=&(g/16\pi^2) m_b \overline{s}_{L} \sigma^{\mu\nu}
            T^a b_{R} G_{\mu\nu}^a\;. \label{eq:o8g}
\end{eqnarray}
 It is the magnetic $\gamma$-penguin operator $O_7$, which plays the crucial
role in this
 decay. The effective Hamiltonian for $b\to d\gamma\gamma$ is
 obtained from Eqs.(\ref{eq:heff}-\ref{eq:o8g}) by the replacement $s\to d$.

In the case of the fourth generation there is an additional
contribution to $b \to  (s,d) \gamma\gamma$ from the virtual
exchange of the fourth generation up quark $t^{'}$. In Fig.1, we
draw the relevant Feynman diagrams which contribute to the decays
$b\to (s,d) \gamma\gamma$. The Wilson coefficients of the dipole
operators are given by
   \begin{equation}\label{ceff4}
      C^{\rm eff}_{7,8}(\mu_b)=C^{\rm (SM)\rm eff}_{7,8}(\mu_b)
      +\frac{V^{*}_{t^{'}q}V_{t^{'}b}}{V^{*}_{tq}V_{tb}}C^{(4)
      {\rm eff}}_{7,8}(\mu_b),
    \end{equation}
where $C^{(4){\rm eff}}_{7,8}(\mu_b)$ represent the contributions
of $t^{'}$ to the Wilson coefficients. We recall here that the CKM
coefficient corresponding to the $t$ quark contribution, i.e.,
$V_{ts}^*V_{tb}$, is factorized in the effective Hamiltonian as
given in Eq. (\ref{Heff_at_mu}). The formulae for calculating the
Wilson coefficients $C_{7,8}^{(4)}(m_W)$ are the same as  in the
SM except exchanging  $t^{'}$ quark  for $t$ quark.

In SM4, the quark mixing matrix can be written as,
\begin{equation}
V = \left (
\begin{array}{lcrr}
V_{ud} & V_{us} & V_{ub} & V_{ub'}\\
V_{cd} & V_{cs} & V_{cb} & V_{cb'}\\
V_{td} & V_{ts} & V_{tb} & V_{tb'}\\
V_{t'd}& V_{t's}& V_{t'b} &V_{t'b'}\\
\end{array} \right ) ,
\end{equation}
where $V_{qb'}$ and $V_{t'q}$ are the elements of the $4\times 4$
CKM mixing matrix
 of the SM4,
 which now contains nine
parameters, i.e., six angles and three phases.  The rest elements
without prime in the matrix are the usual three generation CKM
matrix elements.

The current experimental  bounds of $V^{*}_{t's}V_{t'b}$ and
$V^{*}_{t'd}V_{t'b}$ can be found in ref. \cite{hwj}. The factor
$V^{*}_{t's}V_{t'b}$ can be constrained by the decay $B \to X_s
\gamma$ \cite{huo1,hwj}
\begin{eqnarray}
-11.6\times 10^{-2} < V^{*}_{t's}V^{(1)}_{t'b}< -6.1\times 10^{-2}
,
\end{eqnarray}
or
\begin{eqnarray}
1.9\times 10^{-3} < V^{*}_{t^{'}s}V^{(2)}_{t^{'}b}< 3.6\times
10^{-3}.
\end{eqnarray}
Considering the unitarity conditions of the CKM matrix,
\begin{eqnarray}
 \sum\limits_{i}V_{is}^{*}V_{ib}=0,\,\,\,(i=u,c,t,t'),\label{uni1}
\end{eqnarray}
and taking the average values of the SM $3\times 3$ CKM matrix
elements from Ref. \cite{data}, we obtain,
\begin{eqnarray}
|V^{*}_{t's}V_{t'b}|< 7.6\times 10^{-2}.
\end{eqnarray}
Combining the above constraint,  we arrive at
\begin{eqnarray}
-7.6\times 10^{-2} <
V^{*}_{t's}V^{(1)}_{t'b}< -6.0\times 10^{-2},\\
1.9\times 10^{-3} < V^{*}_{t's}V^{(2)}_{t'b}< 3.6\times 10^{-3}.
\end{eqnarray}
In the following numerical calculation, we take the values
$V^{*}_{t's}V^{(1)}_{t'b}=-7.0\times 10^{-2}$ and $
V^{*}_{t's}V^{(2)}_{t'b}= 2.5\times 10^{-3}$, for illustration.

For the CKM factor $V^{*}_{t'd}V_{t'b}$, we can get its constraint
from the present experimental value of $\Delta M_{B_d}$
\cite{hwj}. Also considering the unitarity conditions of the CKM
matrix,
\begin{eqnarray}
 \sum\limits_{i}V_{id}^{*}V_{ib}=0,\,\,\,(i=u,c,t,t'),
\end{eqnarray}
we finally get
\begin{eqnarray}\label{Vtd}
-1.0\times 10^{-4} < V^{*}_{t'd}V_{t'b}< 0.5\times 10^{-4}.
\end{eqnarray}
In the next section, we will take the values
$V^{*}_{t'd}V_{t'b}=-1.0\times 10^{-4}$, $-0.5\times 10^{-4}$,
$-0.1\times 10^{-4}$, $0.1\times 10^{-4}$ and $  0.5\times
10^{-4}$ in numerical calculation respectively.

\section{The $B_{s,d} \rightarrow \gamma\gamma$ decay rates and phenomenology}

We first calculate the decay $B_s\to \gamma\gamma$ (the formulas
in the case of $B_d\to \gamma\gamma$ can be obtained just by the
replacement of $s\to d$). For this exclusive decay of $B_s$ meson,
we need to deal with both of the long distance and short distance
QCD corrections in the $B_s$ meson side. Applying the
factorization assumption, we shall adapt a phenomenological
approach where the long distance effects are replaced by a few
non-perturbative parameters. In other words we simply evaluate the
hadronic matrix element of $M_{I}+M_{R}$ from one particle
irreducible (1PI) and one particle reducible (1PR) Feynman
diagrams, relying on a phenomenological model. Before doing so, it
is important to note that $M_R$ is apparently non-local  due to
internal $b$ or $s$ quark propagator. To handle these non-local
terms, one observes that the $b$ quark inside the $B_s$ meson
carries most of the meson energy, and its four velocity can be
treated as equal to that of $B_s$.  Hence one may write $b$ quark
momentum as $p=m_bv$ where $v$ is the common four velocity of $b$
and $B_s$. With this parametrization, we have
\begin{eqnarray}
&p\cdot k_1=m_bv\cdot k_1={1\over 2}m_bm_{B_s}=p\cdot k_2,&\nonumber \\
&p'\cdot k_1=(p-k_1-k_2)\cdot k_1=-{1\over2}
m_{B_s}(m_{B_s}-m_b)=p'\cdot k_2&,\label{pp}
\end{eqnarray}
where the second equation is based on a constituent picture
\cite{yao} that $b$ and $\bar{s}$ quarks share the total energy of
$B_s$\footnote{Note that the momentum of $\bar{s}$ quark  is $-p'$
as $p'$ denotes the momentum of $s$ quark in $b\to s
\gamma\gamma$.}. Therefore  $-p'$ should be taken as the four
momentum of the constituent $\bar{s}$ quark. With Eq.(\ref{pp}),
$M_R$ is readily made local. We then compute the amplitude for
$B_s\to \gamma\gamma$ using the following relations
\begin{eqnarray}
\left\langle 0\vert \bar{s}\gamma_{\mu}\gamma_5 b\vert B_s(P)
\right\rangle
&=& -if_{B_s}P_{\mu},\nonumber \\
\left\langle 0\vert \bar{s}\gamma_5 b\vert B_s(P) \right\rangle
&=& if_{B_s}M_B,
\end{eqnarray}
where $f_{B_s}$ is the $B_s$ meson decay constant which is about
$200$ MeV according to recent Lattice QCD calculations
\cite{LATTICE}.

The total amplitude is now separated into a CP-even and a CP-odd
part
\begin{equation}
T(B_s\to \gamma\gamma)=M^+F_{\mu\nu}F^{\mu\nu}
+iM^-F_{\mu\nu}\tilde{F}^{\mu\nu}.
\end{equation}
We find that
\begin{equation}
M^+=-{4{\sqrt 2}\alpha G_F\over 9\pi}f_{B_s}V_{ts}^*V_{tb}\left(
\frac{m_b}{m_{B_s}}B K(m_b^2) +{3C_7\over 8\bar{\Lambda} }\right),
\end{equation}
with $B= -(3C_6+C_5)/4$, $ \bar{\Lambda}=1-m_b/m_{B_s}$, and
\begin{equation}
M^-={4{\sqrt 2}\alpha G_F\over
9\pi}f_{B_s}V_{ts}^*V_{tb}\left(\sum_q
A_qJ(m_q^2)+\frac{m_b}{m_{B_s}}BL(m_b^2)+{3C_7\over 8\bar{\Lambda}
}\right),
\end{equation}
where
\begin{eqnarray}
A_u &=&(C_3-C_5)N_c+(C_4-C_6),\nonumber \\
A_d &=&{1\over 4}\left[(C_3-C_5)N_c+(C_4-C_6)\right],\nonumber \\
A_c &=&(C_1+C_3-C_5)N_c+(C_2+C_4-C_6), \nonumber \\
A_s &=&A_b={1\over 4}\left[(C_3+C_4-C_5)N_c+(C_3+C_4-C_6)\right].
\end{eqnarray}
The functions $J(m^2)$, $K(m^2)$ and $L(m^2)$  are defined by
\begin{eqnarray}
J(m^2)&=&-\frac{1}{2}+\frac{1}{z}I_{00}(m^2),\nonumber \\
K(m^2)&=&-2+\frac{4-z}{z}I_{00}(m^2) ,\nonumber \\
L(m^2)&=&I_{00}(m^2),
\end{eqnarray}
where $z=m_{B_s}^2/m^2$, and
\begin{eqnarray}
I_{00}(m^2)&=&-\int_{0}^{1}{dx} \frac{1}{x}\ln(1-zx+zx^2)\\
&=& \left\{\begin{array}{ll}
2\arctan^2(\sqrt{\frac{z}{4-z}}); & z<4\\
\frac{\pi^2}{2} -2\log^2(\frac{\sqrt{z}+\sqrt{z-4}}{2})
+2i\pi\log(\frac{\sqrt{z}+\sqrt{z-4}}{2}); & z>4
\end{array}\right. .
\end{eqnarray}

The decay width for $B_s\to \gamma\gamma$ is simply
\begin{equation}
\Gamma(B_s\to \gamma\gamma)={m_{B_s}^3\over 16\pi}({\vert M^+\vert
}^2+{\vert M^-\vert }^2).
\end{equation}

To obtain numerical results, we have set light quark masses to
zero and  used \cite{data} $m_t=175 \  {\rm GeV}$, $m_b=5.0 \ {\rm
GeV}$ and $m_c=1.5 \ {\rm GeV}$. Furthermore, we take
$m_{B_s}=5.37 \ {\rm GeV}$, $m_{B_d}=5.28 \ {\rm GeV}$ and
$\alpha=1/ 129$. The numerical values for Wilson coefficients
$C_1-C_8$ evaluated in SM at $\mu=m_b$ are listed in
Table~\ref{t2}.

We first consider the decay mode $B_s\to\gamma\gamma$. Figs.2, 3
and 4 show the branching ratio $Br(B_s\to\gamma\gamma)$ $versus$
the mass of $t'$ and the fourth generation CKM matrix factors
$V^*_{t's}V_{t'b}^{(1)}$ and $V^*_{t's}V_{t'b}^{(2)}$,
respectively.  From these  figures, we find that the SM4
predictions for the branching ratio $Br(B_s\to\gamma\gamma)$ are
rather different from that of the standard model. The decay rates
can be enhanced by about 1-2 orders relative to the SM prediction
in the reasonable range of $m_{t'}$ and $V^*_{t's}V_{t'b}$. This
implies that the  new physics signal of  the fourth generation
Model may come from the decay of $B_s\to\gamma\gamma$. The future
experiment of LHCb in CERN can test it. From Fig.2, we find that
the branching ratio of $B_s\to\gamma\gamma$ increase with the mass
of $t'$ for the fourth generation CKM matrix factor,
$V^*_{t's}V_{t'b}=-7.0\times 10^{-2}$ and $2.5\times 10^{-3}$.
When $V^*_{t's}V_{t'b} =-7.0\times 10^{-2}$, the curve increase
rapidly with $m_{t'}$. But when $V^*_{t's}V_{t'b} =2.5\times
10^{-3}$, the  branching ratio changes very slowly. Because the
fourth generation CKM matrix factor, $V^*_{t's}V_{t'b}$, is small
in the latter case.
 It can be seen from   Eq.~(\ref{ceff4}) that  the new physics contribution is
proportional to the size of the CKM factor $V^*_{t's}V_{t'b}$,
thus it is very important in the new physics contribution with
$m_{t'}$. From Figs.3 and 4, one can also see   that,
   the branching ratio of $B_s\to\gamma\gamma$ decease
 with $V^*_{t's}V_{t'b}^{(1)}$ but increase with
 $V^*_{t's}V_{t'b}^{(2)}$,
since   $V^*_{t's}V_{t'b}^{(1)}$ is negative and gives the
opposite contributions as SM to the coefficients $C^{eff}_{7,8}$
(see Eq. (\ref{ceff4}) ).

For the decay $B_d\to\gamma\gamma$, it is easy to find from Figs.5
and 6 that the new physics contribution to the branching ratio is
tiny. It is shown that it is not suitable to test the new physics
signals from the 4th generation in this channel. The prediction of
the branching ratio in the 4th generation model is   between $2.35
\times 10^{-8}$ and $2.45\times 10^{-8}$, when we take the
reasonable values of $m_{t'}$ (from 200 GeV to 1000 GeV) and
$V^*_{t'd}V_{t'b}$ (see Eq. (\ref{Vtd})). This numerical result is
fully in the range of the SM prediction (see Eq. (4)).
  The reason  is  that the experimental constraint of the 4th generation CKM
matrix
 element factor $V^*_{t'd}V_{t'b}$ is very stringent (only
at $10^{-4}$ level, see Eq.(\ref{Vtd})). We also find that the
curves change little along with the variations of $m_{t'}$ and
$V^*_{t'd}V_{t'b}$. The
 reason is also from the tiny $V^*_{t'd}V_{t'b}$.  We can see that the branching
ratios of both
 $B_s\to\gamma\gamma$ and $B_s\to\gamma\gamma$ decays strongly depend on the
size of the related
 mixing elements.

\section{Conclusion}

As a summary, the size of new physics contributions to the rare
decays of $B_{s,d}\to \gamma\gamma$ from the fourth generation
strongly depends on the values of the masses of the new heavy up
quark $t'$ and the 4th generation CKM matrix elements,
$V^*_{t's}V_{t'b}$ and$V^*_{t'd}V_{t'b}$.  The results in the 4th
generation model  are quite different from the SM case for decay
$B_s\to\gamma\gamma$, its branching ratio can be enhanced by about
1-2 orders of magnitude. A large branching ratio of  $B_{s}\to
\gamma \gamma$ decay measured in experiment may
 be a new physics signal of the 4th generation model.
  But for decay
$B_d\to\gamma\gamma$, because of the tiny values of 4th generation
CKM matrix element, $V_{t'd}^* V_{t'b}$, the new physics
contribution to ${\rm BR}(B_{d}\to\gamma\gamma)$ is very small and
can not be distinct from the SM prediction.

\vspace{1cm}

\noindent {\bf ACKNOWLEDGMENT}

W.J. Huo acknowledges the support  from the Chinese Postdoctoral
Science Foundation.from the Natural Sciences and Engineering
Research Council of Canada. C.D. L\"u acknowledges the support  by
National  Science Foundation of China under Grants No.~90103013
and 10135060. And Z.J.~Xiao acknowledges the support by the
National Natural Science Foundation of China under Grants
No.~10075013 and 10275035, and by the Research Foundation of
Nanjing Normal University under Grant No.~214080A916.

\newpage

\begin{table}[htbp]
\begin{center}
\caption{ The values of Wilson coefficients $C_1-C_8$ at $\mu
=m_b$ in SM} \label{t2} \vspace{.1cm}
\begin{tabular}{cccc cccc} \hline
 $C_1$ & $C_2$ & $C_3$ & $C_4$ & $C_5$ & $C_6$ & $C^{eff}_7$ (SM)&
$C^{eff}_8$(SM) \\ \hline $0.235$ & $-1.100$ & $-0.011$ & $0.024 $
&$-0.007$ &$0.029$ & $0.306 $ & $0.146$ \\ \hline
\end{tabular}
\end{center}
\end{table}

\newpage

\begin{figure}[h]
\vspace{40pt}
\begin{minipage}[]{\textwidth}
    \epsfig{file=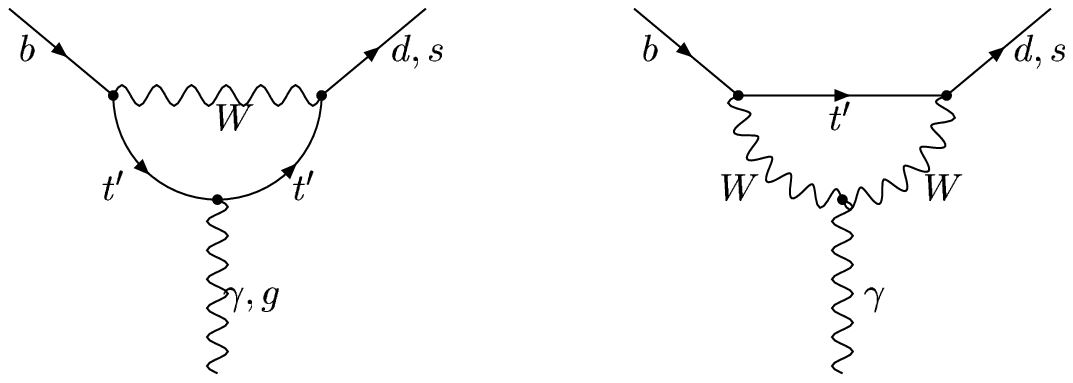,bbllx=2cm,bblly=10.5cm,bburx=15cm,bbury=25cm,%
width=12cm,angle=0} \vspace{-180pt} \caption{Magnetic photon and
gluon penguin diagrams  with the fourth generation $t'$ quark.}
\label{fig:fig1}
\end{minipage}
\end{figure}

\newpage
\begin{figure}[ht]
\centerline{
    \epsfig{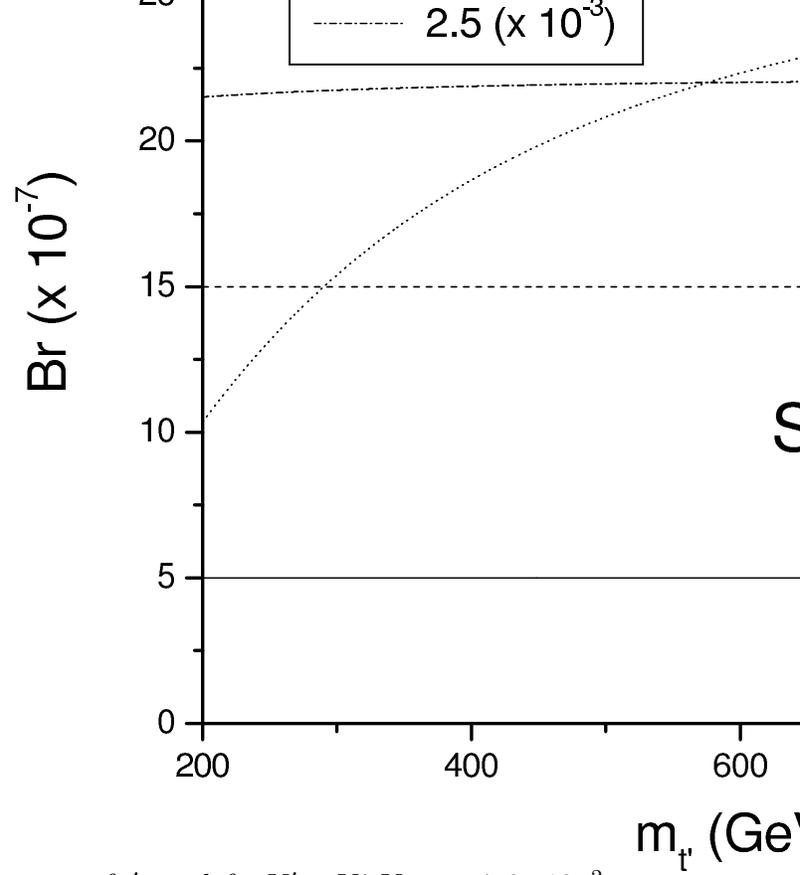}  }
\caption{Branching ratios of $B_s\to\gamma\gamma$ against  the
mass of $t'$ quark for $V'=V^*_{t's}V_{t'b}=-7.0\times 10^{-2}$
and $2.5\times 10^{-3}$ respectively. The lower two horizontal
lines represent the SM prediction the branching ratio.}
\label{fig:fig2}
\end{figure}

\newpage
\begin{figure}[htb]
\vspace{0pt}
\centerline{
    \epsfig{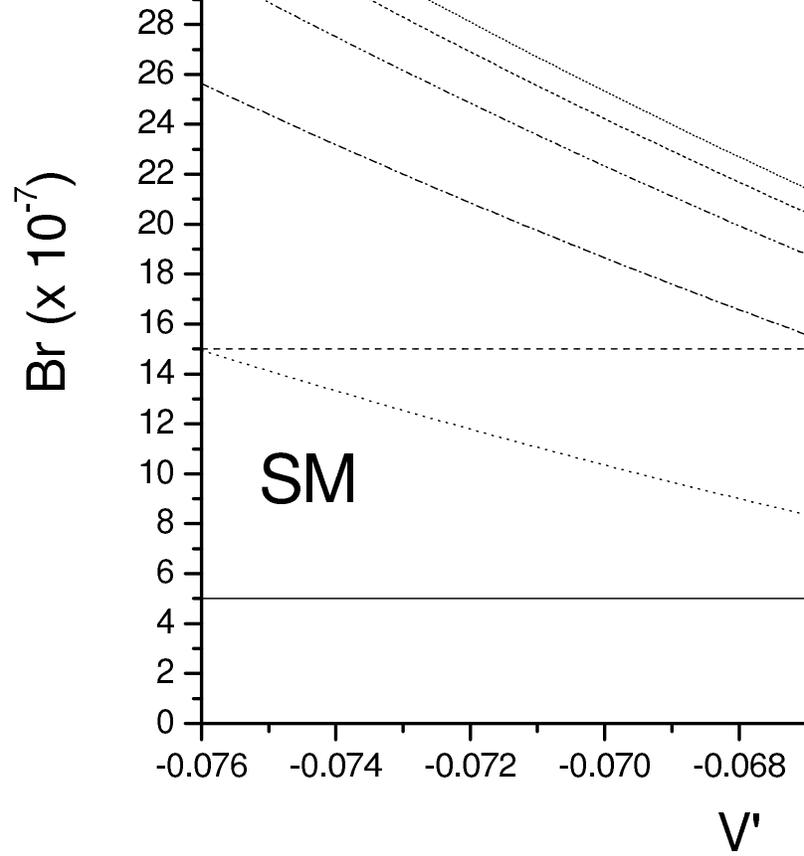}  } \vspace{0pt} \caption{Branching ratios of
$B_s\to\gamma\gamma$ as a function of $V'=V^*_{t's}V_{t'b}^{(1)}$
for $m_{t'}= $ 200 GeV, 400 GeV, 600 GeV, 800 GeV and 1000 GeV,
respectively. The  two horizontal lines represent the SM
prediction the branching ratio.}
 \label{fig:fig3}
\end{figure}

\newpage

\begin{figure}[htb]
\centerline{
    \epsfig{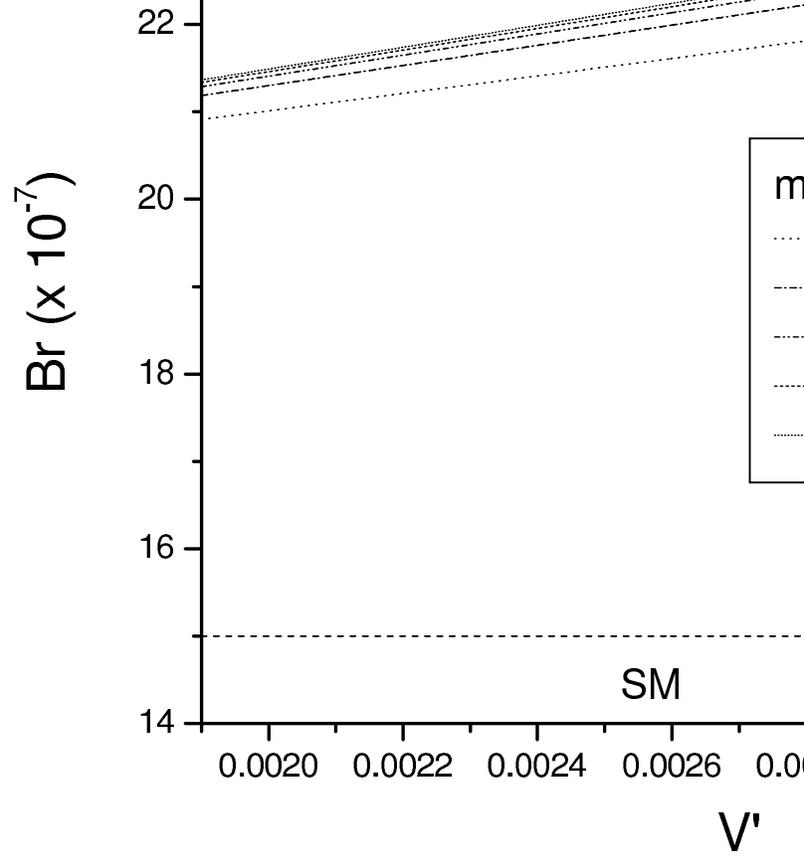}  } \vspace{0pt} \caption{Branching ratios of
$B_s\to\gamma\gamma$ as a function of  $V'=V^*_{t's}V_{t'b}^{(2)}$
for $m_{t'}= $ 200 GeV, 400 GeV, 600 GeV, 800 GeV and 1000 GeV,
respectively.}\label{fig:fig4}
\end{figure}

\newpage

\begin{figure}[htb]
\centerline{
    \epsfig{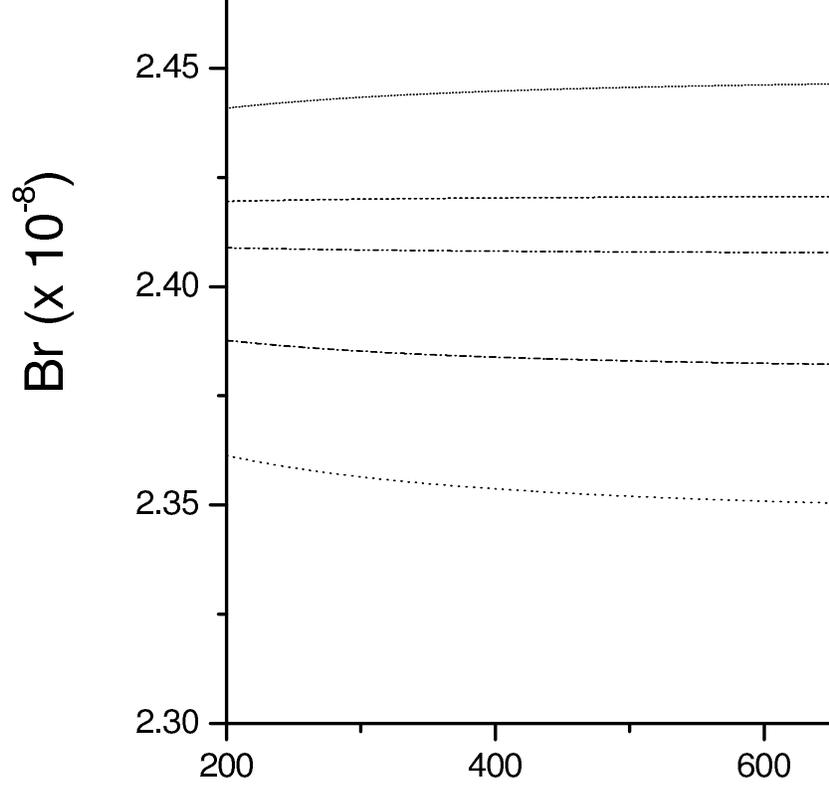}  } \vspace{-20pt} \caption{Branching ratios of
$B_d\to\gamma\gamma$ as a function of the mass of $t'$ quark for
$V'=V^*_{t'd}V_{t'b}$ takes several values in the range of
$-1.0\times 10^{-4}$ and $0.5\times 10^{-4}$.}\label{fig:fig5}
\end{figure}
\newpage

\begin{figure}[htb]
\centerline{
    \epsfig{file=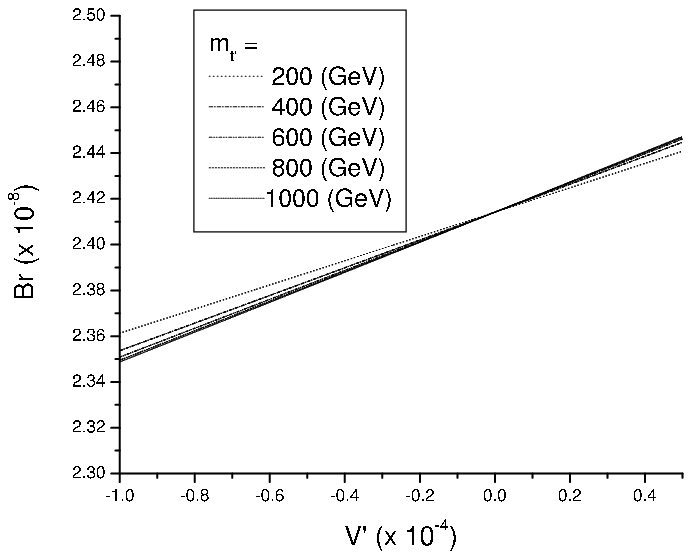,bbllx=-3cm,bblly=19cm,bburx=5cm,bbury=25cm,%
width=12cm,angle=0}  }
 \vspace{-20pt} \caption{Branching ratios of $B_s\to\gamma\gamma$
 as a function of  $V'=V^*_{t'd}V_{t'b}$ for $m_{t'}= $ 200 GeV, 400 GeV,
 600 GeV, 800 GeV and 1000 GeV, respectively.}\label{fig:fig6}
\end{figure}

\end{document}